\documentclass[reprint,superscriptaddress,nofootinbib,prd,floatfix]{revtex4}
\pdfoutput=1
\usepackage{graphicx}
\usepackage[dvipsnames]{xcolor}
\usepackage{amsmath}
\allowdisplaybreaks
\usepackage{amssymb}

\usepackage{slashed}
\usepackage[utf8]{inputenc}
\usepackage[colorlinks=true,linkcolor=blue,bookmarksopen,bookmarksnumbered]{hyperref}
\usepackage{color} 

\usepackage[normalem]{ulem}
\usepackage{soul}
\usepackage{units}
\usepackage{rotating}
\usepackage{hhline,multirow,tabularx}
\usepackage{bm}

\usepackage{cleveref} 
\usepackage{nameref} 
\usepackage[export]{adjustbox}
\usepackage{mdframed}
\usepackage{comment}
\usepackage{natbib}
\usepackage{aliascnt} 

\usepackage{graphicx}
\usepackage{subcaption}

\def\bea#1\eea{\begin{align}#1\end{align}}

\usepackage{lipsum}
\usepackage{hyperref}
\begin{document}

\title{Studying the impact of background field on coupling constants in EiC and EicC}

\author{Cong Li}
\affiliation{School of Information Engineering, Zhejiang Ocean University, Zhoushan, 316022, Zhejiang, China}
\author{Yutong Xiao}
\affiliation{School of Information Engineering, Zhejiang Ocean University, Zhoushan, 316022, Zhejiang, China}


\begin{abstract}

We studied the influence of background fields on coupling constants in various kinematic regions. We then evaluated these effects through the Bethe-Heitler process in electron-nucleus collisions at the EIC and EicC.

\end{abstract}
\maketitle

\section{Introduction}

In the context of Color Glass Condensate (CGC) \cite{CGC1,CGC2,CGC3} and Quark-Gluon Plasma (QGP) \cite{QGP1,QGP2,QGP3,QGP4,QGP5}, certain quantum chromodynamics (QCD) processes, which are typically perturbative, can exhibit non-perturbative behavior in experimental observations. This suggests that, under high-density gluon fields, the nature of these interactions undergoes significant changes. High-density gluon fields notably affect interactions by altering the coupling constants of strong interactions. We aim primarily to quantify the impact of background fields on these coupling constants, a challenge that we previously addressed in our earlier work \cite{Li1}. This paper investigates the influence of background fields across different kinematic regions on coupling constants. The implications of this effect are substantial, since most studies related to strong interactions, owing to color confinement, cannot be separated from nuclei or hadrons. Thus, the effects of gluon background fields on coupling constants are critical. Moreover, we seek to resolve a long-standing question: do the interacting particles exchanged during the process originate from charges or from the vacuum? Of course, the prevailing belief remains that they derive from the vacuum.

In this paper, we propose to investigate the impact of background fields on coupling constants in electron-nucleus (eA) collisions. While this involves a Quantum Electrodynamics (QED) process, it remains focused on our central theme, the influence of background fields on coupling constants. In Quantum Chromodynamics (QCD), gluon fields and the strong coupling constant have corresponding analogs in QED: photon fields and the electromagnetic coupling constant, respectively. The analysis of QED processes offers several advantages, including their simplicity and a relatively clean background field. Additionally, the parameterization of photon distribution functions associated with the photon background field has been extensively studied \cite{EPA1,EPA2} and measured \cite{star}, providing a stable and well-understood platform for examining the effects. 

The establishment of the Electron-Ion Collider (EIC) in the United States and the Electron-Ion Collider in China (EicC) enables experimental measurement of this effect. Our numerical evaluations are based on the Bethe-Heitler (BH) process in eA collisions at the EIC and EicC. The peripheral photon field of the nucleus  acts as the venue for interactions, where electron radiation occurs within this background photon field, influencing the corresponding coupling constant. Furthermore, this peripheral photon field supplies a stable and well-defined incoming photon beam, which is integral to the process. Clearly, both the EIC and EicC are well-suited environments for studying this process, primarily because nuclei play multiple roles, facilitating both theoretical calculations and experimental measurements.

This study also builds on another previous study \cite{cc1,cc2,cc3}, which treated the peripheral photon field of the nucleus as a background field and assessed the effects of Coulomb scattering when electrons traverse this field. Based on it, in the current study, we will also consider Coulomb corrections and Sudakov factors, the latter representing effects from final-state soft-photon radiation. Moreover, we will investigate how the background field influences azimuthal asymmetries through its impact on the coupling constant.

The structure of this paper is organized as follows. In the next section, we will briefly introduce the distribution function of nucleus peripheral photons, accounting for Coulomb corrections, and discuss quantifying the influence of background fields on coupling constants. The third section will present numerical estimates of the impact of corrected coupling constants on the BH process in eA collisions at the EIC and EicC, including impacts on cross section and azimuthal asymmetry. Finally, we will conclude with a succinct summary that highlights the significant implications of our findings for understanding the interplay between background fields and coupling constants in high-energy collisions.

\section{Correction of coupling constant by background field}


When studying the Bethe-Heitler process in electron-nucleus (eA) collisions at the EIC and EicC, the cross section can be expressed using Transverse Momentum Distribution (TMD) factorization, representing the convolution of the hard part and the photon TMD distributions. As electrons traverse the photon field, they experience multiple scatterings within the background, collectively referred to as Coulomb corrections. This process can occur either prior to emitting a photon or after one has been radiated in BH process. Coulomb corrections can be incorporated into the correlation function of photons in the form of a closed-loop gauge link, which can be expressed as,
\begin{equation}
\int \frac{dy^- d^2 y_\perp}{P^+ (2\pi)^3} e^{ik \cdot y} \langle A | F_{(+\perp)}^\mu(0) U^\dagger(0_\perp) U(y_\perp) F_{(+\perp)}^\nu(y) | A \rangle \bigg|_{(y^+=0)} = \frac{\delta_\perp^{\mu\nu}}{2} x f_1^\gamma(x, k_\perp^2) + \left( \frac{k_\perp^\mu k_\perp^\nu}{k_\perp^2} - \frac{\delta_\perp^{\mu\nu}}{2} \right) x h_1^{\perp \gamma}(x, k_\perp^2)
\end{equation}
where \(\delta_\perp^{\mu\nu} = -g^{\mu\nu} + p^\mu n^\nu + p^\nu n^\mu\), resulting in \(k_\perp^2 = \delta_\perp^{\mu\nu} k_{\perp \mu} k_{\perp \nu}\). The functions \(f_1^\gamma(x, k_\perp^2)\) and \(h_1^{\perp \gamma}(x, k_\perp^2)\) represent the unpolarized and linearly polarized photon TMDs, respectively. The terms \(U^\dagger(0_\perp)U(y_\perp)\) comprise the transverse gauge link, effectively integrating the Coulomb correction of the hard part into the photon TMD distributions. The specific form of the gauge link is given by.
\begin{equation}
U(y_\perp) = \mathcal{P} \exp\left( ie \int_{-\infty}^{+\infty} dz^- A^+(z^-, y_\perp) \right)
\end{equation}
It is crucial to note that, since photons do not carry electric charge, the correlation function does not require gauge links to maintain gauge invariance. The gauge potential can be expressed as,
\begin{equation}
V(y_\perp) \equiv e \int_{-\infty}^{+\infty} dz^- A^+(z^-, y_\perp) = \frac{\alpha Z}{\pi} \int d^2 q_\perp \frac{e^{-iy_\perp \cdot q_\perp} F(q_\perp^2)}{q_\perp^2 + \delta^2}
\end{equation}
where \(\delta\) denotes the mass of the photon, serving to regulate infrared divergences, and \(F\) is the nuclear charge form factor. Our numerical evaluations utilize the nuclear charge form factor obtained from the STARlight Monte Carlo generator \cite{starlight},
\begin{equation}
F(|\mathbf{k}|) = \frac{4\pi \rho^0}{|\mathbf{k}|^3 A} \left[\sin(|\mathbf{k}|R_A) - |\mathbf{k}|R_A \cos(|\mathbf{k}|R_A)\right] \frac{1}{a^2 |\mathbf{k}|^2 + 1}
\end{equation}
where \(R_A = 1.1 A^{1/3} \, \text{fm}\) and \(a = 0.7 \, \text{fm}\), closely resembling the Woods-Saxon distribution. The field strength in the correlation function is expressed as,
\begin{equation}
\mathcal{F}^\mu(x,y_\perp) \equiv \int_{-\infty}^{+\infty} dy^- e^{ix P^+ y^-} F_{(+\perp)}^\mu(y^-, y_\perp) = \frac{Ze}{4\pi^2} \int d^2 q_\perp e^{-iy_\perp \cdot q_\perp} \frac{(iq_\perp^\mu) F(q_\perp^2 + x^2 M_p^2)}{q_\perp^2 + x^2 M_p^2}
\end{equation}
where \(x\) represents the longitudinal momentum fraction carried by the photon, and \(M_p\) denotes the mass of the proton. Thus, the photon correlation function can be rewritten as,
\begin{equation}
\int \frac{d^2 y_\perp d^2 y_\perp'}{4\pi^3} e^{ik_\perp \cdot (y_\perp - y_\perp')} \mathcal{F}^\mu(x, y_\perp) \mathcal{F}^{*\nu}(x, y_\perp') e^{i[V(y_\perp) - V(y_\perp')]} = \frac{\delta_\perp^{\mu\nu}}{2} x f_1^\gamma(x,k_\perp^2) + \left( \frac{k_\perp^\mu k_\perp^\nu}{k_\perp^2} - \frac{\delta_\perp^{\mu\nu}}{2} \right) x h_1^{\perp \gamma}(x,k_\perp^2)
\end{equation}
The required photon TMDs can be determined through orthogonality.
\begin{equation}
x f_1^\gamma(x, k_\perp^2) = \int \frac{d^2 y_\perp d^2 y_\perp'}{4\pi^3} e^{ik_\perp \cdot (y_\perp - y_\perp')} \mathcal{F}_\mu(x, y_\perp) \mathcal{F}^{*}_{\nu}(x, y_\perp') \delta_\perp^{\mu\nu} e^{i[V(y_\perp) - V(y_\perp')]}
\end{equation}
\begin{equation}
x h_1^{\perp \gamma}(x, k_\perp^2) = \int \frac{d^2 y_\perp d^2 y_\perp'}{4\pi^3} e^{ik_\perp \cdot (y_\perp - y_\perp')} \mathcal{F}_\mu(x, y_\perp) \mathcal{F}^{*}_{\nu}(x, y_\perp') \left( \frac{2k_\perp^\mu k_\perp^\nu}{k_\perp^2} - \delta_\perp^{\mu\nu}\right) e^{i[V(y_\perp) - V(y_\perp')]}
\end{equation}

The propagator of the electron in the presence of the photon background field will be influenced by the background, leading to Coulomb corrections. Simultaneously, the photon propagator in the photon background field will also be affected. We start with the Wick contraction of the photon propagator in the background photon field,
\begin{equation}
\langle B | A_\mu(x) A_\nu(y) | B \rangle = \int \frac{d^4 k}{(2\pi)^4} e^{-ik \cdot (x-y)} \frac{-ig_{\mu\nu}}{k^2 + i\epsilon} |\langle B | k \rangle|^2
\end{equation}
where \(B\) denotes the background photon field. The propagator in the background differs from that in vacuum; it convolves the vacuum propagator with a background field photon distribution function \(|\langle B | k \rangle|^2 = f(k)\), representing the probability density of on-shell photons with momentum \(k\) in the background field. The influence on the photon propagator can be included in an effective coupling constant,
\begin{equation}
\alpha_{\text{eff}}(k) = \alpha_e f(k)
\end{equation}
where \(\alpha_e\) is the electromagnetic coupling constant, approximately \(1/137\). The effective coupling \(\alpha_{\text{eff}}\) reflects the actual measured electromagnetic strength within this background field. A high density of photons in the background enhances the effective coupling constant, even more so resulting in a transition from perturbative to non-perturbative behavior. In simpler terms, examining the same physical process across different kinematic regions results in variations in $k$, which ultimately lead to a change in \(\alpha_{\text{eff}}(k)\). To evaluate the enhancement of the effective coupling constant due to the photon distribution function across different kinematic intervals, we apply the Mean Value Theorem for Integrals to get the average value of $f(k)$ in the  kinematic intervals.
\begin{equation}
\bar{f}(k) = \frac{\int_a^b dk_z \int_c^d k_\perp dk_\perp f(k)}{\int_a^b dk_z \int_c^d k_\perp dk_\perp}
\end{equation}
The expression denotes the average value of the distribution function within the interval \(k_z \in [a,b]\) and \(k_\perp \in [c,d]\). When \(\bar{f}(k) > 1\), it indicates that, within the kinematic interval, the background field enhances the coupling constant. Conversely, if \(\bar{f}(k) < 1\), it suggests a diminishing effect. A value of \(\bar{f}(k) > 137\) implies that perturbation theory may not be applicable in that kinematic range.

The photon distribution function outside the atomic nucleus can be described using the equivalent photon approximation (EPA) \cite{EPA1,EPA2,EPA3,EPA4},
\begin{equation}
n(k_z, k_\perp^2) = \frac{Z^2 \alpha_e}{\pi^2} \frac{k_\perp^2}{|k_z|} \left[\frac{F(k_\perp^2 + k_z^2)}{(k_\perp^2 + k_z^2)}\right]^2
\end{equation}
where \(Z\) is the nuclear charge number, and \(F\) is the nuclear charge form factor. It is crucial to distinguish between \(n(k_z, k_\perp^2)\) and \(f(k)\).
\begin{equation}
f(k) = n(k_z, k_\perp^2) 2E_k (2\pi)^3
\end{equation}
In the Bethe-Heitler process, an electron absorbs a photon from the atomic nucleus and emits a photon. During the emission process, occurring within the background photon field, the coupling constant is affected by the background field, transforming the original \(\alpha_e\) into \(\alpha_{\text{eff}}(k)\).

\section{Observables}

As mentioned above, we investigate the impact of the background field on the coupling constant through the BH process,
\begin{equation}
e(\bar{P}) + \gamma(xP + k_\perp) \rightarrow \gamma(p_1) + e(p_2).
\end{equation}
where $xP+k_\perp$ is the total momentum transfer through multiple photon exchanges. We focus on the kinematic region associated with the correlation limit. In this region, the total transverse momentum $q_\perp=k_\perp = p_{1\perp} + p_{2\perp} \approx 0$ is very small, thus $p_{1\perp} \approx -p_{2\perp} \approx P_\perp$. 

Due to the well-separated scales involved, we consider the contributions of large logarithmic terms arising from soft photon radiation. The resummation of all large logarithm terms results in the Sudakov factor $\exp\left(-\frac{\alpha_e}{2\pi} \ln^2\left(\frac{P_\perp^2}{\mu_r^2}\right)\right)$ with $\mu_r = \frac{2e^{-\gamma_E}}{|r_\perp|}$. After convoluting the Sudakov factor over the collision parameter space, the differential cross section becomes,
\begin{equation}
\frac{d\sigma}{dy_\gamma d^2 P_\perp d^2 q_\perp} = H_{\text{Born}} \int \frac{d^2 r_\perp}{(2\pi)^2} e^{ir_\perp \cdot q_\perp} e^{-\frac{\alpha_e}{2\pi} \ln^2\left(\frac{P_\perp^2}{\mu_r^2}\right)} \int d^2 k_\perp e^{ir_\perp \cdot k_\perp} x f_1^\gamma(x, k_\perp^2),
\end{equation}
where $y_\gamma$ is the rapidity of the emitted photon. The coefficient of the hard part can be readily obtained from the Ref. \cite{cc3,crosssection1,crosssection2},
\begin{equation}
H_{\text{Born}} = 2\alpha_e^2 f(y_\gamma,P_\perp^2) z^2 \frac{1 + (1 - z)^2}{P_\perp^4},
\end{equation}
where $z$ is the longitudinal momentum fraction of the incoming electron carried by the outgoing photon. At the energy of the EIC and EicC, the mass of the electron is negligible. With the above materials, we can numerically evaluate the effect on the cross section. 

\begin{figure}[htbp]
    \centering
    \begin{subfigure}{0.5\textwidth}
        \centering
        \includegraphics[width=\textwidth]{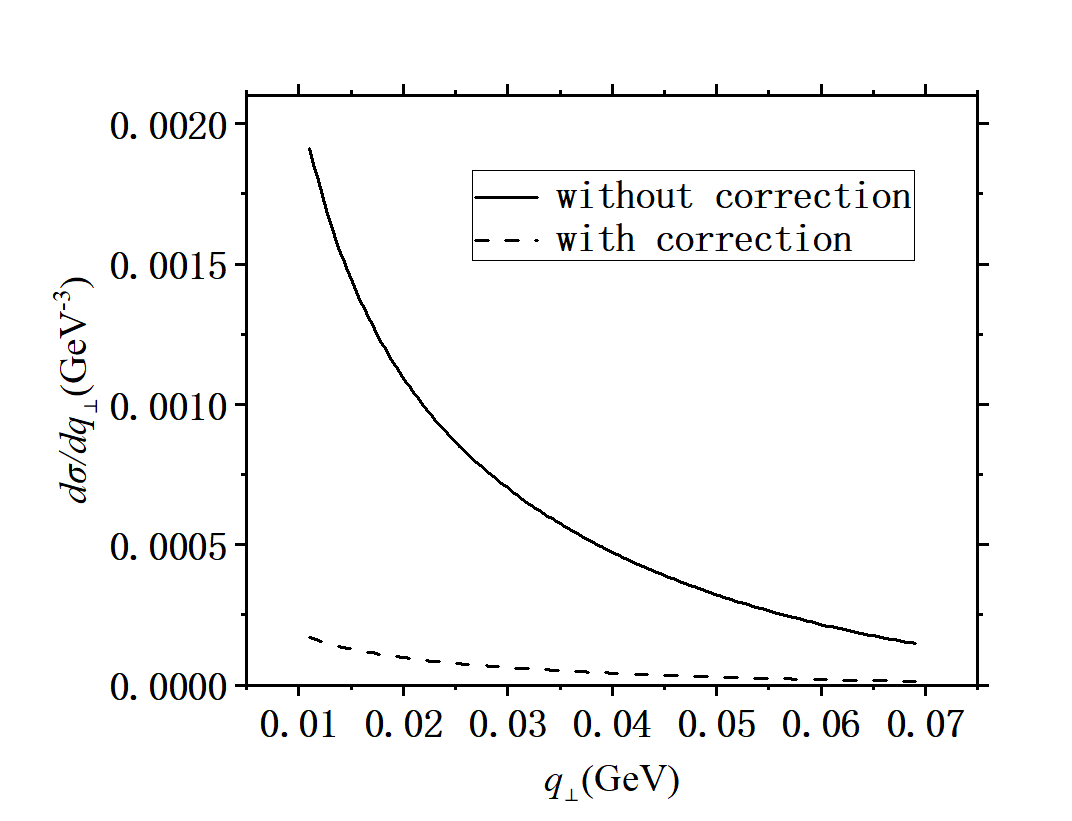} 
    \end{subfigure}\hfill
    \begin{subfigure}{0.5\textwidth}
        \centering
        \includegraphics[width=\textwidth]{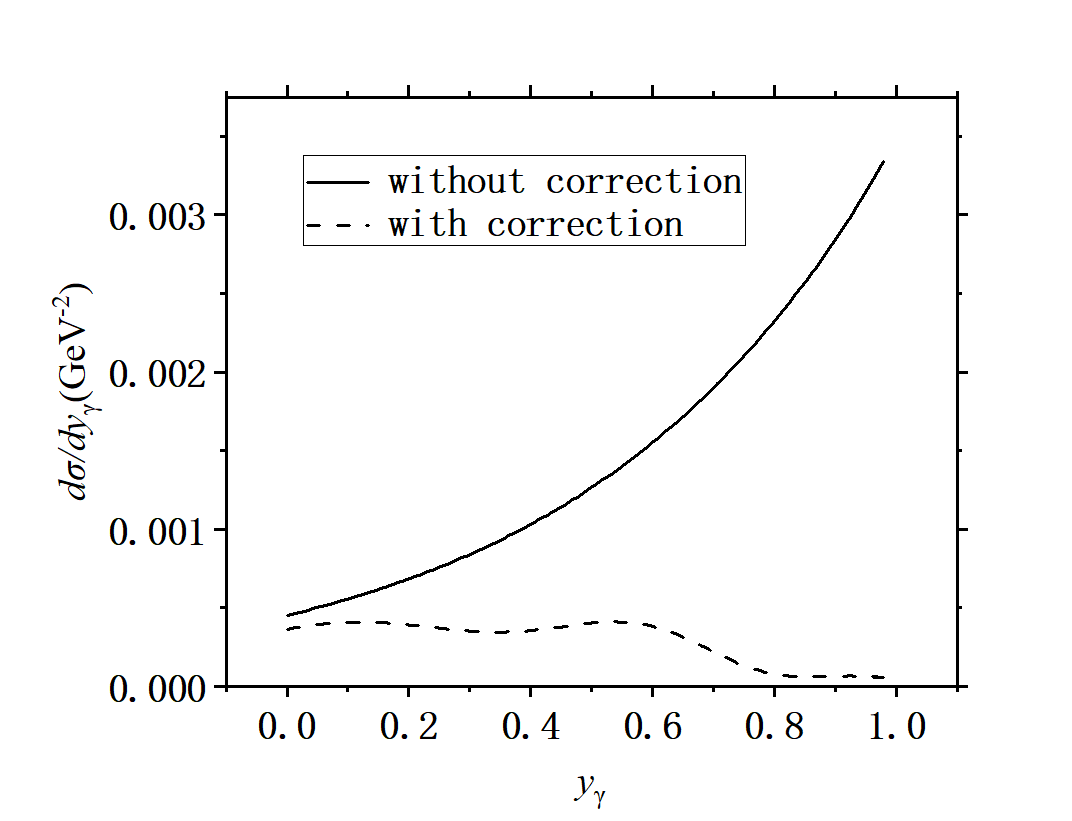} 
    \end{subfigure}
    \caption{The cross section as the function of $q_{\perp}$ (left panel) and $y_\gamma$ (right panel) for a Pb target at EIC. $P_{\perp}$ is integrated over the regions [ $300 \mathrm{MeV}, 400 \mathrm{MeV}$ ]. In the left plot, the emitted photon rapidity $y_\gamma$ is integrated over $[0.5,1]$. In the right plot, the total transverse momentum $q_{\perp}$ is fixed to be 20 MeV .}
\end{figure}
\begin{figure}[htbp]
    \centering
    \begin{subfigure}{0.5\textwidth}
        \centering
        \includegraphics[width=\textwidth]{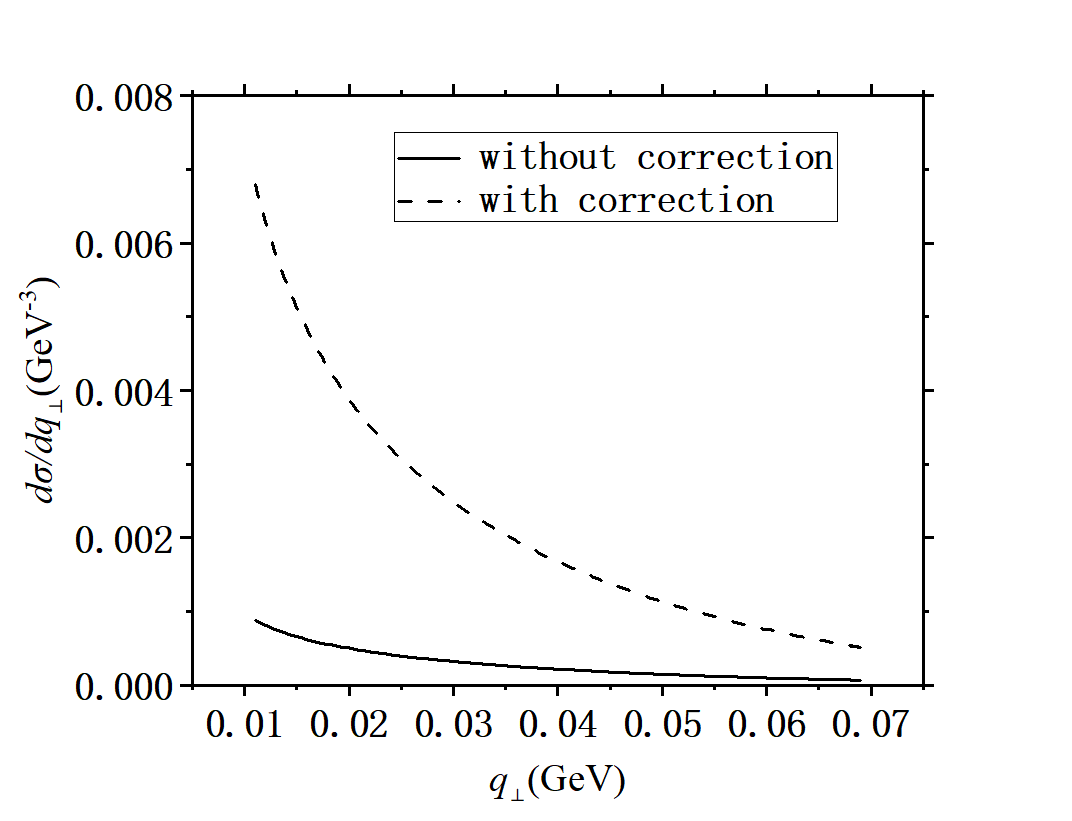} 
    \end{subfigure}\hfill
    \begin{subfigure}{0.5\textwidth}
        \centering
        \includegraphics[width=\textwidth]{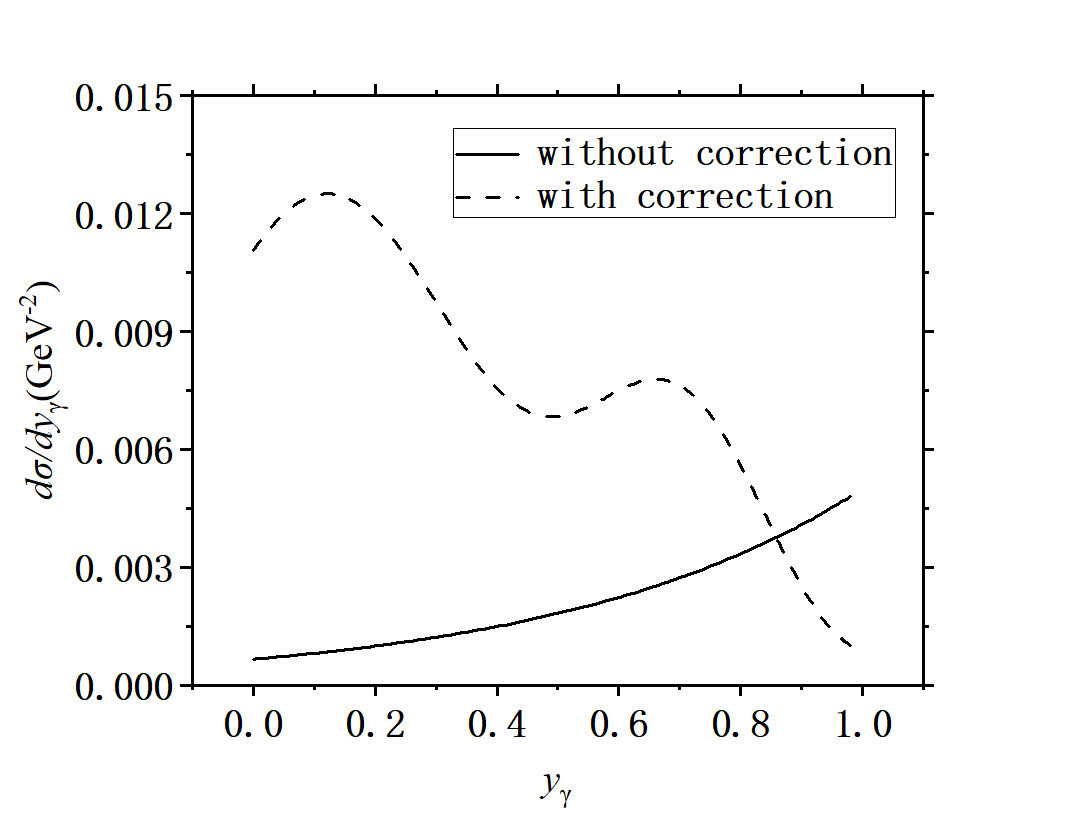} 
    \end{subfigure}
    \caption{The cross section as the function of $q_{\perp}$ (left panel) and $y_\gamma$ (right panel) for a Pb target at EIC. $P_{\perp}$ is integrated over the regions [ $200 \mathrm{MeV}, 300 \mathrm{MeV}$ ]. In the left plot, the emitted photon rapidity $y_\gamma$ is integrated over $[0.1,0.5]$. In the right plot, the total transverse momentum $q_{\perp}$ is fixed to be 20 MeV .}
\end{figure}
\begin{figure}[htbp]
    \centering
    \begin{subfigure}{0.5\textwidth}
        \centering
        \includegraphics[width=\textwidth]{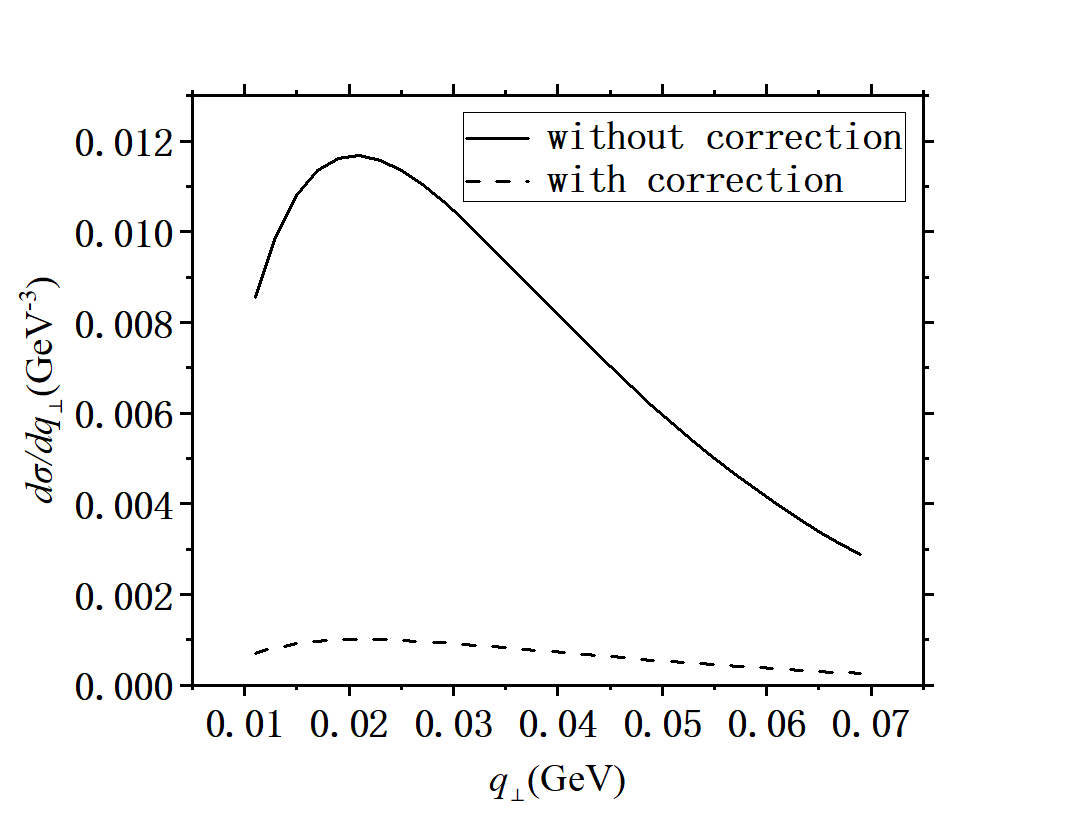} 
    \end{subfigure}\hfill
    \begin{subfigure}{0.5\textwidth}
        \centering
        \includegraphics[width=\textwidth]{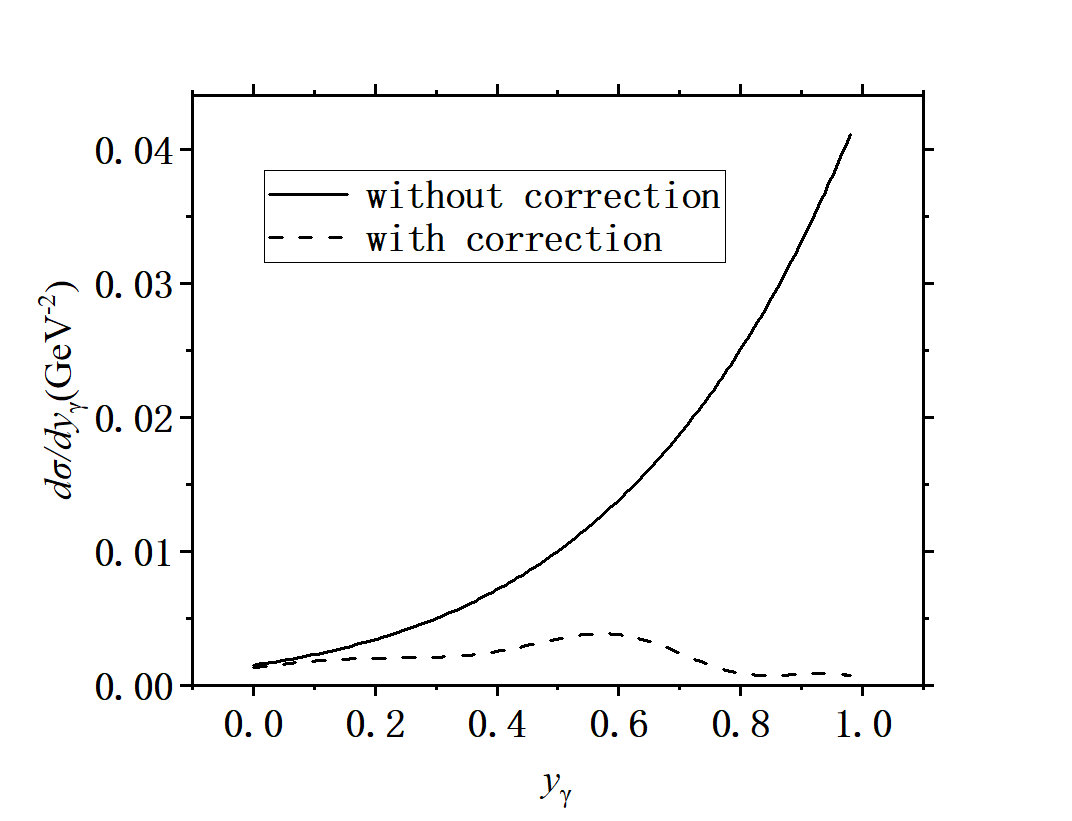} 
    \end{subfigure}
    \caption{The cross section as the function of $q_{\perp}$ (left panel) and $y_\gamma$ (right panel) for a Pb target at EicC. $P_{\perp}$ is integrated over the regions [ $300 \mathrm{MeV}, 400 \mathrm{MeV}$ ]. In the left plot, the emitted photon rapidity $y_\gamma$ is integrated over $[0.5,1]$. In the right plot, the total transverse momentum $q_{\perp}$ is fixed to be 20 MeV .}
\end{figure}
\begin{figure}[htbp]
    \centering
    \begin{subfigure}{0.5\textwidth}
        \centering
        \includegraphics[width=\textwidth]{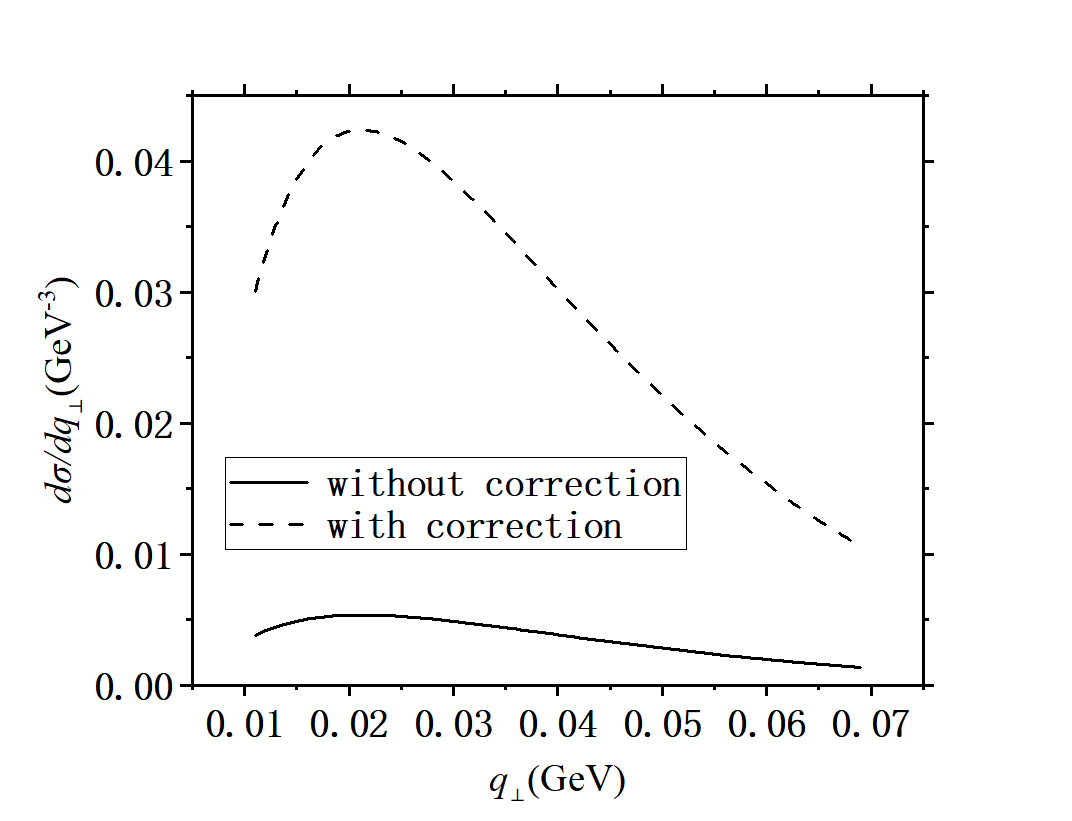} 
    \end{subfigure}\hfill
    \begin{subfigure}{0.5\textwidth}
        \centering
        \includegraphics[width=\textwidth]{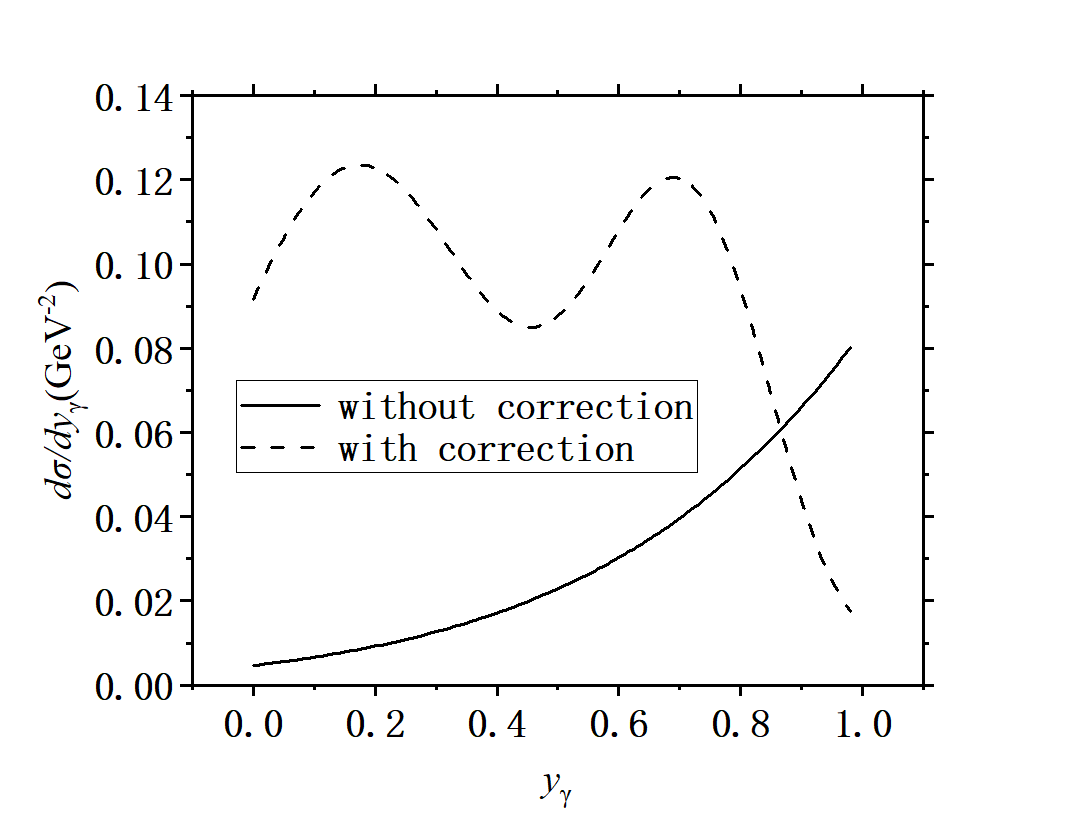} 
    \end{subfigure}
    \caption{The cross section as the function of $q_{\perp}$ (left panel) and $y_\gamma$ (right panel) for a Pb target at EicC. $P_{\perp}$ is integrated over the regions [ $200 \mathrm{MeV}, 300 \mathrm{MeV}$ ]. In the left plot, the emitted photon rapidity $y_\gamma$ is integrated over $[0.1,0.5]$. In the right plot, the total transverse momentum $q_{\perp}$ is fixed to be 20 MeV .}
\end{figure}
We assess the cross section's dependence on the rapidity of the final-state photon, as well as the total transverse momentum at the energies of EIC and EicC, where the electron beam and heavy ion beam energies are 18 GeV and 100 GeV for EIC, and 3.5 GeV and 8 GeV for EicC, respectively. For comparison, we also evaluate the differential cross section without this correction. To showcase the effect of varying kinematic regions on the amplitude, we evaluate the cross sections for $\bar{f}(k)= 0.4$ and 97.9, indicating the weakening and enhancement of the coupling constant due to the background field in the respective kinematic regions $P_\perp \in [0.3, 0.4]$ and $y_\gamma \in [0.5, 1]$, as well as $P_\perp \in [0.2, 0.3]$ and $y_\gamma \in [0.1, 0.5]$. The effect on the amplitude is shown to be significant in Figure 1-4. Firstly, the effect greatly enhances or weakens the cross section that initially did not consider this impact. The general relation between weakening or enhancement aligns with the comparison of $\bar{f}(k)$ to 1. However, no clear systematic pattern in magnitude can be detected. Nevertheless, it is certain that in high photon density kinematic regions, the probability of photon emission increases, resulting in a larger cross section, and vice versa. Secondly, it is very interesting that the dependence of the corrected cross section on rapidity is very strange, it is oscillating, which should be affected by EPA. It is also convergent, and the same EPA is also convergent. If experimental measurements observe this effect, it may aid in accurately understanding the coupling constant in the background field and confirm a previous belief: bosons exchanged during the interaction come from the vacuum itself, rather than from charge radiation. 

We also evaluated the effect on azimuthal asymmetries. In the BH process, the incoming photon originates from the nucleus, and thus it is transversely polarized. This causes the angle between the final-state particles to no longer be isotropic, allowing the observable $ \cos(2\phi) $ azimuthal asymmetry if a virtual photon instead of a real one is emitted in the final state. From the Ref.\cite{crosssection1,crosssection2,cc3}, the azimuthal-dependent cross section can be easily obtained,
\begin{equation}
    \begin{aligned}
        \frac{d\sigma}{dy_\gamma d^2 P_\perp d^2 q_\perp} = \int \frac{d^2 r_\perp}{(2\pi)^2} &e^{ir_\perp \cdot q_\perp}e^{-\frac{\alpha_e}{2\pi} \ln^2(Q^2/\mu_r^2)}\\& \times \int d^2 k_\perp e^{ir_\perp \cdot k_\perp} \left\{ H_{\text{Born}}' x f_1^\gamma(x, k_\perp^2) + H_{\text{Born}}^{\cos(2\phi)} [2(\hat{k}_\perp \cdot \hat{P}_\perp)^2 - 1] x h_1^{\perp\gamma}(x, k_\perp^2) \right\}
    \end{aligned}
\end{equation}
where $\hat{k}_\perp = \frac{k_\perp}{|k_\perp|}$ and $\hat{P}_\perp = \frac{P_\perp}{|P_\perp|}$ are unit transverse vectors. The modified hard parts are as follows.
\begin{equation}
H_{\text{Born}}' = 2\alpha_e^2 f(y_\gamma,P_\perp^2) z^2 \left[ \frac{1 + (1 - z)^2}{(P_\perp^2 + (1 - z) Q^2)^2} - \frac{2Q^2 P_\perp^2 z^2 (1 - z)}{(P_\perp^2 + (1 - z) Q^2)^4} \right]
\end{equation}
\begin{equation}
H_{\text{Born}}^{\cos(2\phi)} = 2\alpha_e^2 f(y_\gamma,P_\perp^2) z^2 \left[-\frac{2Q^2 P_\perp^2 z^2 (1 - z)}{(P_\perp^2 + (1 - z) Q^2)^4}\right]
\end{equation}
Finally, we evaluate the azimuthal asymmetries at the energy of the EIC. The average value of $\langle \cos(2\phi) \rangle$ is given by,
\begin{equation}
\langle \cos(2\phi) \rangle = \frac{\int \frac{d\sigma}{dP.S.} \cos(2\phi) dP.S.}{\int \frac{d\sigma}{dP.S.} dP.S.}
\end{equation}
where $dP.S.$ is the phase space factor. As shown in Figure 5, the rapidity dependence of azimuthal asymmetries is minimally affected by this effect, increasing with rapidity. However, the momentum dependence of the asymmetries is significantly influenced, with the asymmetry reduced by a factor of three before and after correction. All in all, the trends of both dependencies remain consistent, presumably due to the ratio of $H_{\text{Born}}'$ to $H_{\text{Born}}^{\cos(2\phi)}$ partially canceling the effect. We have not displayed the azimuthal asymmetries at EicC energy, as the asymmetries are comparatively small at this energy, especially in this kinematic range where further suppression by the effect would occur. In general, observing the effect through azimuthal asymmetries presents some challenges, since the asymmetry is approximately 3\% if present.
\begin{figure}[htbp]
    \centering
    \begin{subfigure}{0.5\textwidth}
        \centering
        \includegraphics[width=\textwidth]{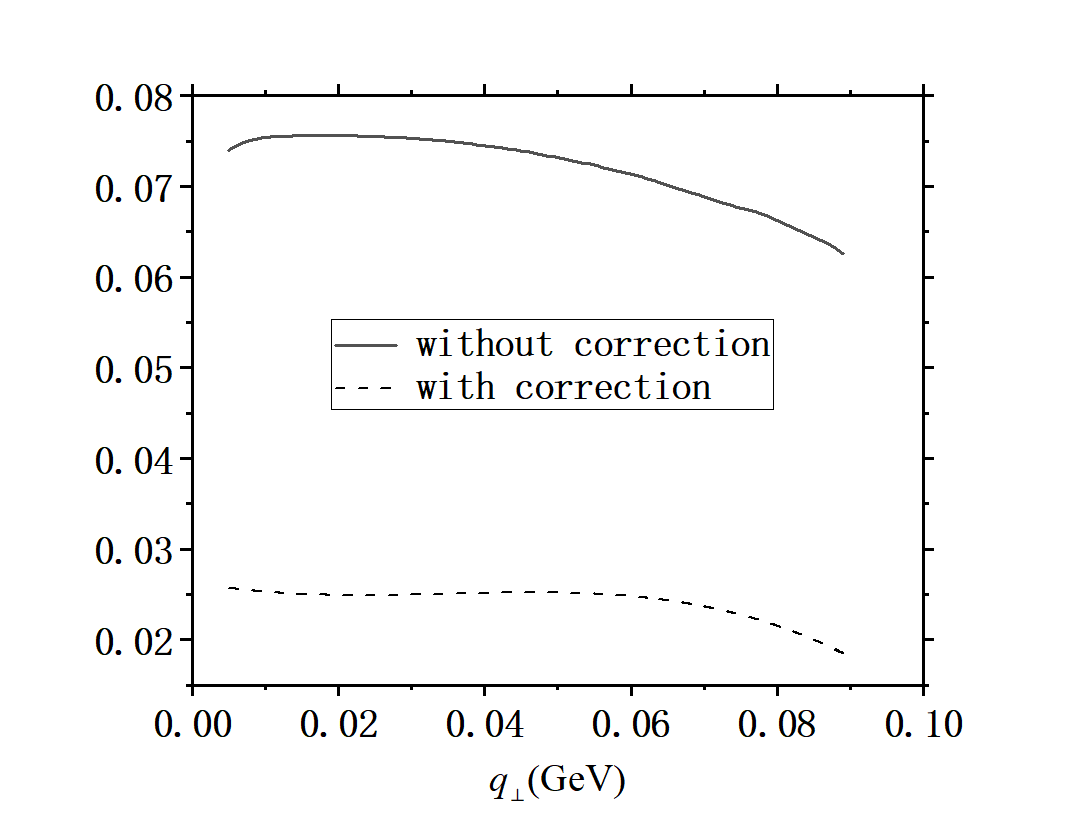} 
    \end{subfigure}\hfill
    \begin{subfigure}{0.5\textwidth}
        \centering
        \includegraphics[width=\textwidth]{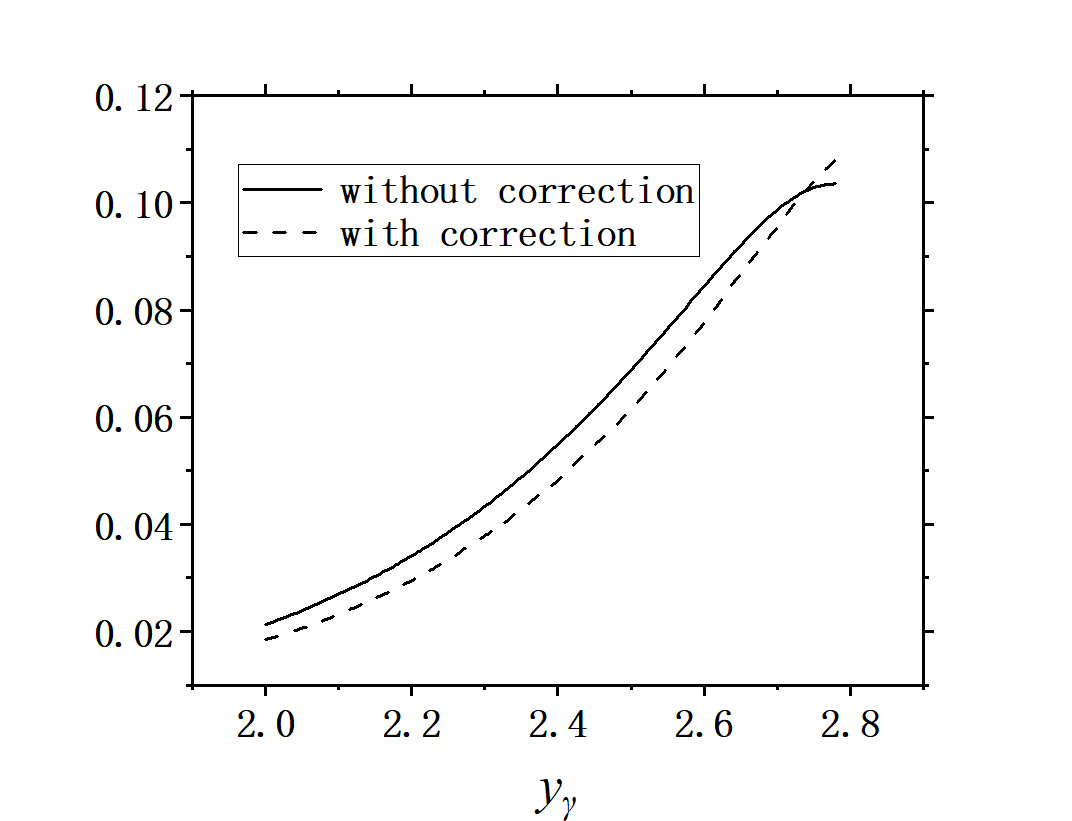} 
    \end{subfigure}
    \caption{Fig. 3. The azimuthal asymmetry as the function of $q_{\perp}$ (left panel) and $y_\gamma$ (right panel) with and without taking into account the corrections of background field on coupling constant for a Pb target at EIC. $Q^2$ is fixed to be $Q^2=4 \mathrm{GeV}^2$. The $P_{\perp}$ is integrated over the  region $[1.5 \mathrm{GeV}, 2 \mathrm{GeV}]$. In the left plot, the emitted photon rapidity $y_\gamma$ is integrated over the region $[2,2.8]$. In the right plot, the total transverse momentum $q_{\perp}$ is fixed to be 50 MeV .}
\end{figure}

\section{Summary}

In this paper, we propose to study the impact of the background field on the coupling constant through the Bethe-Heitler process in electron-nucleus (eA) collisions at the EIC and EicC. The influence of the background field on the coupling constant is inherently complex, and our investigation begins with the impact of the background field on photon propagators. This effect resembles the Coulomb correction, which originated from the modification of the electron propagator due to multiple scattering of electrons in the background field. The presence of the atomic nucleus alters the distribution of photons in the background field, ultimately leading to this effect.

In the latter part of the paper, we numerically assess the influence of this effect at the energies of EIC and EicC, considering its magnitude across different kinematic regions. In kinematic regions with high photon density, the coupling constant increases anomalously, resulting in a larger cross section. In contrast, in another region, the cross section decreases. In our estimates of azimuthal asymmetries, the effect is not markedly pronounced, which we hypothesize arises from cancellation during the ratio calculations. We anticipate that observing this effect will contribute to more precise theoretical predictions within QCD, which currently lacks high precision. Studies of this effect also help to confirm that the interacting force-carrying particles come from the background field, although this was previously widely believed.

\section*{Acknowledgments}
AI tools were used to improve sentence fluency and make English more natural.

\bibliography{ref}

\end{document}